\documentstyle[12pt,epsf]{article}
\def\det{\hbox{\rm det}}

\def\Tr{{\rm Tr}}

\def\l{{\cal L}}

\def\S{{\cal S}}

\def\Z{{\bf Z}}
\def\N{{\bf N}}

\def\half{\frac{1}{2}}
\def\RR{R-R }
\def\NSNS{NS-NS }

\def\s{\bf s}

\def\pmb#1{\setbox0=\hbox{#1}
 \kern-.025em\copy0\kern-\wd0
 \kern.05em\copy0\kern-\wd0
 \kern-.025em\raise.0433em\box0 }

\def\3{\ss}
\def\sq{\hbox{\rlap{$\sqcap$}$\sqcup$}}
\def\qed{\ifmmode\sq\else{\unskip\nobreak\hfil
\penalty50\hskip1em\null\nobreak\hfil\sq
\parfillskip=0pt\finalhyphendemerits=0\endgraf}\fi}

\newcommand{\ket}[1]{|#1\rangle}
\newcommand{\bra}[1]{\langle#1|}

\def\mm {\advance\voffset by -1.5cm
\advance\hoffset by -2.1cm
\textwidth=17.3cm
\textheight=20.0cm}

\def\b,#1,#2(#3){\left|B#1,#2\right>_{#3}}
\def\e#1{e^{ik^{#1}X}}
\def\gso#1,#2{\frac{1}{4}(1#1(-1)^F)(1#2(-1)^{{\tilde F}})}

\def\xxx#1           {{\sf hep-th/#1} }

\def\npb#1(#2)#3     {Nucl. Phys. {\bf B#1} (#2) #3 }
\def\rep#1(#2)#3     {Phys. Rept.{\bf #1} (#2) #3 }
\def\pla#1(#2)#3     {Phys. Lett. {\bf #1A} (#2) #3 }   
\def\plb#1(#2)#3     {Phys. Lett. {\bf #1B} (#2) #3 }
\def\prl#1(#2)#3     {Phys. Rev. Lett.{\bf #1} (#2) #3 }
\def\prd#1(#2)#3     {Phys. Rev. {\bf D#1} (#2) #3 }
\def\ap#1(#2)#3      {Ann. Phys. {\bf #1} (#2) #3 }
\def\rmp#1(#2)#3     {Rev. Mod. Phys. {\bf #1} (#2) #3 }
\def\cmp#1(#2)#3     {Comm. Math. Phys. {\bf #1} (#2) #3 }
\def\mpla#1(#2)#3    {Mod. Phys. Lett. {\bf A#1} (#2) #3 }
\def\ijmp#1(#2)#3    {Int. J. Mod. Phys. {\bf A#1} (#2) #3 }
\def\cqg#1(#2)#3     {Class. Quant. Grav. {\bf #1} (#2) #3 }
\def\am#1(#2)#3      {Adv. Math. {\bf #1} (#2) #3 }
\def\im#1(#2)#3      {Invent. Math. {\bf #1} (#2) #3 }
\def\jhep#1(#2)#3    {JHEP {\bf #1}(#2) #3 }
\def\npps#1(#2)#3    {Nucl.Phys.Proc.Suppl. {\bf #1}(#2) #3 }
\def\jgp#1(#2)#3     {J. Geom. Phys. {\bf #1}(#2) #3 }

\def\dstyler#1       {\documentstyle{report}[#1]}
\def\dstylea#1       {\documentstyle{article}[#1]}
\def\bd              {\begin{document}}
\def\ed              {\end{document}}
\def\be	             {\begin{equation}}
\def\ee              {\end{equation}}
\def\ba              {\begin{eqnarray}}
\def\ea              {\end{eqnarray}}
\def\ni              {\noindent}
\def\bb#1            {\begin{thebibliography}{#1}}
\def\eb              {\end{thebibliography}}

\def\etal {{\em et al.} }
\def\w    {\;_\wedge}
\def\ie   {{\em i.e.}}
\def\ibid {{\em ibid.}}

\advance\voffset by -1.5cm
\advance\hoffset by -2.1cm
\def\hu{u_{\mbox{\scriptsize c}}}
\def\hv{v_{\mbox{\scriptsize c}}}
\def\htau{\tau_{\mbox{\scriptsize c}}}
\def\uo{u_{\mbox{\scriptsize o}}}
\def\vo{v_{\mbox{\scriptsize o}}}
\newcommand{\rmd}{{\rm d}}

\def\cf{{\em cf. }}
\def\G{\Gamma}
\def\D{\Delta}
\def\L{\Lambda}
\def\S{\Sigma}
\def\a{\alpha}
\def\b{\beta}
\def\g{\gamma}
\def\d{\delta}
\def\e{\varepsilon}
\def\x{{\bf x}}
\def\m{\mu}
\def\n{\nu}
\def\s{\sigma}
\def\r{\rho}
\def\l{\lambda}
\def\t{\tau}
\def\o{\omega}
\def\v{\varrho}
\def\mc{\mathcal}
\def\N{\nabla}
\def\p{\partial}
\newcommand{\bz}{{\bar{z}}}
\newcommand{\bw}{{\bar{w}}}
\newcommand{\la}{\langle}
\newcommand{\ra}{\rangle}

\begin{document}

\thispagestyle{empty}
\def\thefootnote{\fnsymbol{footnote}}
\begin{flushright}
  hep-th/0105238 \\
  AEI-2001-054
\end{flushright}
\vskip 0.5cm

\begin{center}\LARGE
{\bf Boundary Superstring Field Theory Annulus Partition Function in the Presence of Tachyons}
\end{center}
\vskip 1.0cm
\begin{center}
{\large G. Arutyunov\footnote{E-mail address: {\tt agleb@aei.mpg.de}}
\footnote{On leave of absence from Steklov Mathematical Institute,
Gubkin str.8, 117966, Moscow, Russia}, 
A. Pankiewicz\footnote{E-mail address: {\tt apankie@aei.mpg.de}},
B. Stefa\'nski, jr.\footnote{E-mail address: {\tt bogdan@aei.mpg.de}}}

\vskip 0.5cm
{\it Max-Planck-Institut f\"ur Gravitationsphysik, Albert-Einstein Institut \\ 
Am M\"uhlenberg 1, D-14476 Golm, Germany}
\end{center}

\vskip 1.0cm

\begin{center}
May 2001
\end{center}

\vskip 1.0cm

\begin{abstract}
\noindent
We compute the Boundary Superstring Field Theory partition function on the annulus
in the presence of independent linear tachyon profiles on the two boundaries. 
The \RR sector is found to contribute non-trivially to the derivative terms of the space-time 
effective
action. In the process we 
construct a boundary state description of D-branes in the presence of a linear tachyon. We quantize
the open string in a tachyonic background 
and address the question of open/closed string duality.
\end{abstract}

\vfill
\setcounter{footnote}{0}
\def\thefootnote{\arabic{footnote}}
\newpage

\renewcommand{\theequation}{\thesection.\arabic{equation}}

\section{Introduction}

In the last few years significant progress has been made in understanding 
the open string tachyon dynamics. It has become clear that open string 
tachyon condensation describes the decay of unstable
D-branes into stable ones or into the closed string vacuum. 
Initially the discussion was based on the first quantized
string theory \cite{Sen,WittenK,BG}. 
Subsequently tachyon condensation has been investigated with 
a remarkable degree of accuracy in Cubic Open String Field Theory~\cite{Wittenoc}
by using the level truncation approximation~\cite{LT}.
From the world-sheet point of view the tachyon condensation process
is then viewed as the RG flow relating conformal field theories with Neumann 
and Dirichlet boundary conditions~\cite{HKM,DD}.
More recently it has been argued~\cite{GS,KMM1,Ghoshal,KMM2,Andreev,gsnov,KL} that    
the Boundary String Field Theory (BSFT)~\cite{WittenBSFT,Sch} also 
provides a suitable description of the tachyon condensation. In particular 
the exact tree level tachyon potential, the ratios of the brane tensions~\cite{GS,KMM1} 
as well as the low-energy effective 
action for massless fluctuations around a tachyonic soliton~\cite{AFTT}  
are obtained quite naturally in this setting. 

A basic object of the BSFT is an effective space-time action $S_{\mbox{\scriptsize eff}}$ 
considered as a functional 
of the open string background fields. In the supersymmetric case, with
which we will be mainly concerned here, 
the BSFT action is known~\cite{KMM2,tsey00,Mar,Niar} 
to coincide with the partition function $Z$
of the open string boundary sigma model~\cite{FT}.  
In the sigma model approach the Weyl and diffeomorphism 
invariant string action on a world-sheet $\Sigma$ with boundaries
is modified by including boundary perturbations which correspond to turning on space-time 
background fields. For example one may turn on a background gauge field $A_{\mu}$ by including a
boundary perturbation\footnote{In the introduction we restrict ourselves to the 
bosonic part of the boundary action.}
\be
S_{\mbox{\scriptsize gauge}}=-i\int_{\p\S}\rmd s \, A_{\mu}(X)
\frac{\partial}{\partial s}X^{\mu}\, ,
\ee
This perturbation is marginal as it preserves both Weyl 
and diffeomorphism invariance. In the study of the open superstring background tachyon one
includes a boundary perturbation
\be
S_{\mbox{\scriptsize tachyon}}=\frac{1}{2}\int_{\p\S} \rmd s \, T^2(X)\,,
\label{tachpert}
\ee
where $T(X)$ is a tachyon profile. A particularly simple case is the linear profile 
$
T(X)=u_\mu X^\mu\,,
$
for some non-negative constants $u_\mu$ and $\rmd s$ is the diffeomorphism invariant 
length element on the boundary $\p\S$ of the world-sheet. In this case the boundary sigma model 
is exactly solvable. The higher open string background fields 
can also be incorporated in this approach~\cite{Frolov} though the corresponding action 
defines a non-renormalizable theory. 

One important part of the investigation of the supersymmetric boundary sigma model is a study 
of contributions of higher genus Riemann surfaces to $Z$. This should give new insight into  
the effective action $S_{\mbox{\scriptsize eff}}$, provided that the correspondence 
$S_{\mbox{\scriptsize eff}}=Z$
holds for higher Riemann surfaces as well. In this paper we consider 
the case of the annulus, a subject already 
being under discussion in the current 
literature \cite{RVY}-\cite{Bardakci}. 

In general one defines (as in the case of Polyakov's string~\cite{Polyakov}) the path integral
to include an integration over world-sheet metrics, modulo the symmetries of the theory. 
In particular the sigma model  action is still invariant with respect 
to a subgroup of the diffeomorphism group compatible
with the boundary conditions. The boundary action~(\ref{tachpert}) is not Weyl-invariant, 
while the
bulk action is. This too should be accounted for in a path integral
formulation. Since such a path integral formalism, properly accounting for the remaining gauge 
symmetry, is not presently available,  
the first problem to study is to consider some fixed metric 
$g_{\alpha\beta}$ on the annulus and to compute $Z[g_{\alpha\beta}]$, the path  integral 
over the $X^\mu$'s. 

In this paper we consider the simplest case of the flat metric on the annulus 
for which the bosonic part of the action is (we set $\a'=2$)
\ba
S&=&\frac{1}{8\pi}\int_a^1\int_0^{2\pi}r\;\rmd r\rmd\phi\left(\frac{1}{r^2}\bigl(\p_\phi X(r,\phi)
\bigr)^2+\bigl(\p_rX(r,\phi)\bigr)^2\right)\nonumber\\
&&+\frac{1}{8\pi}
u_{\mu}u_{\nu}\int_0^{2\pi}\rmd\phi X^{\mu}X^{\nu}(1,\phi)+\frac{1}{8\pi}av_{\mu}v_{\nu}
\int_0^{2\pi}\rmd\phi X^{\mu}X^{\nu}(a,\phi)\,,\label{bosact}
\ea
where $u\,,v$ are tachyons on the $r=1$ and $r=a\le 1$ boundaries of the annulus, respectively; 
the presence of $a$ 
in the last term of the action is due to the diffeomorphism invariant measure $\rmd s$. 
A more detailed discussion of the
action involving fermions will be given in Section~\ref{sec1a}. 

The computation of the annulus partition function can be done in several ways
and below we briefly summarise those used in this paper.  
Perhaps the most 
direct approach is via the Green's function method, discussed in Section~\ref{sec2}.
The classical field configurations minimizing the bosonic action~(\ref{bosact})
are subjected to the following boundary conditions\footnote{Our boundary 
conditions differ from the ones used in~\cite{RVY,VY,Bardakci}. They do 
agree with~\cite{Suy} and~\cite{Alisha} who also use the diffeomorphism invariant measure.}
\be
\partial_rX^\mu(1,\phi)+u_\mu u_\nu X^\nu(1,\phi)=0\,,\qquad 
-\partial_rX^\mu(a,\phi)+v_\mu v_\nu X^\nu(a,\phi)=0\, 
\label{greenbc}
\ee
and computing the path integral by the Green's function approach one may 
conveniently choose the Green's functions to obey the same type of 
boundary conditions.
The two limiting cases $u=0\,,\infty$ correspond to Neumann and Dirichlet boundary 
conditions, respectively. 

In Section~\ref{sec2} we derive the boundary conditions for fermions, taking into 
account the different spin structures on the annulus. In particular we show that 
the choice of spin structure is equivalent to identifying 
the boundary fermion, treated as a Lagrangian multiplier, 
in terms of the bulk fermions. In Section~\ref{sec2} we construct the Green's
functions both in the \NSNS and \RR sectors for different spin structures 
and use them to derive the corresponding contributions to the 
partition function.\footnote{The Green's functions in the \NSNS 
sector of the boundary sigma model were considered in \cite{VY} and \cite{Alisha} but 
only for one particular choice of the spin structure.}  

In Section~\ref{sec3} we discuss another way to compute the partition function. We map the 
annulus to the cylinder via
\be
\tau_{\mbox{\scriptsize c}}=\ln r\,,\qquad \sigma_{\mbox{\scriptsize c}}=\phi\,,
\ee
and construct boundary states~\cite{CLNY-1,PC} 
$\ket{B,u}\,,\bra{B,v}$ corresponding to the absorption and 
emission of closed strings from the D-branes with background tachyons
turned on.\footnote{Boundary states for bosonic D-branes and the \NSNS sector have been 
independently constructed in~\cite{da,Bardakci}.}
Specialising to a tachyon profile with $u_{\m}$ and $v_{\m}$ in the same direction (say $\m=9$) 
the bosonic part of the boundary states (in that direction) satisfy
\be
\left(\partial_{\tau_{\mbox{\scriptsize c}}} X^9(1,\sigma_{\mbox{\scriptsize c}})
+u^2 X^9(1,\sigma_{\mbox{\scriptsize c}})\right)\ket{B,u,0}=0\,,\qquad
\bra{B,v,-l}\left(-\partial_{\tau_{\mbox{\scriptsize c}}} X^9(l,\sigma_{\mbox{\scriptsize c}})
+av^2 X^9(l,\sigma_{\mbox{\scriptsize c}})\right)=0\,,
\label{bdrbc}
\ee
with $l=-\ln a$. The partition function on the cylinder is
\be
Z_{\mbox{\scriptsize cylinder}}(u,v)=\int_0^\infty\rmd l\bra{B,v}e^{-l H_c}\ket{B,u}\,,
\ee
where $H_c$ is the (conformal) closed string Hamiltonian. 
In this formalism the disc partition function is
\be
Z_{\mbox{\scriptsize disc}}(u)=\left<\mbox{vacuum}\right.\ket{B,u}\,,
\ee
which can be shown to coincide with the one computed in~\cite{GS,KMM1,KMM2,WittenBSFT}. 
The boundary state  approach emphasizes the 
conformal nature of the 
bulk theory - in the bulk
the usual Virasoro generators $L_n,{\tilde L}_n$ are well defined and satisfy the standard 
algebra.
For a conformal boundary perturbation the boundary states satisfy \cite{CLNY}
\be
(L_n-{\tilde L}_{-n})\ket{B}=0\,,
\ee 
indicating that Weyl invariance is not broken on the boundary. The $\ket{B,u}$ satisfy no such 
simple relation. 

The construction of the boundary states involves finding suitably normalized 
coherent states which satisfy~(\ref{bdrbc}) and fermionic boundary conditions in the \NSNS and 
\RR sectors. These states have to be
invariant under the closed string GSO projection. Due to the presence of fermionic zero-modes 
this places a restriction on the allowed 
boundary states in the \RR sector~\cite{BG}.  
In particular it is well known that with no 
background tachyon the
D$p$-brane \RR boundary state is GSO invariant for $p$ even/odd in Type II A/B, respectively. 
So for example there is no GSO invariant \RR boundary state corresponding to the non-BPS D9-brane of 
Type IIA. We show that in the presence of a non-zero background tachyon in one direction 
(say $u^9\neq 0$) a GSO invariant non-BPS D9-brane \RR boundary state does exist. 
The normalization of this state
depends linearly on $u^9$ and so becomes zero in the non-BPS D9-brane limit. In the limit 
$u^9\to\infty$ it reduces to the boundary state of the BPS D8-brane. 

Finally, we change coordinates on the annulus again, taking the world-sheet time to
be periodic
\be
\sigma_{\mbox{\scriptsize o}}=-\tau_{\mbox{\scriptsize c}}\frac{\pi}{l}\,,\qquad
\tau_{\mbox{\scriptsize o}}=\sigma_{\mbox{\scriptsize c}}\frac{\pi}{l}\,,
\ee
and compute the functional integral on the annulus as an open string 
partition function. In this case the boundary conditions are
\be
\partial_{\sigma_{\mbox{\scriptsize o}}} X^9(\tau_{\mbox{\scriptsize o}},0)
-\frac{u^2 l}{\pi} X^9(\tau_{\mbox{\scriptsize o}},0)=0\,,\qquad
\partial_{\sigma_{\mbox{\scriptsize o}}} X^9(\tau_{\mbox{\scriptsize o}},\pi)
+\frac{av^2 l}{\pi} X^9(\tau_{\mbox{\scriptsize o}},\pi)=0\,.
\label{openbc}
\ee
In the first part of Section~\ref{sec4} the open string on a strip with 
boundary conditions relevant to tachyonic perturbations is analysed.
The system is canonically quantised and found to have a countably infinite spectrum. Our 
partition function is found to factorise on closed string poles (with residues depending on the
tachyons) giving it the interpretation of a 
transition amplitude for a closed string propagating between two non-BPS D9-branes in the presence of 
background tachyons.
We compute a renormalised, tachyon dependent normal-ordering constant of the
open string Hamiltonian which we expect to be compatible with the open/closed string duality.
The path integral on the annulus is then computed as an open string partition 
function with boundary conditions~(\ref{openbc}) in the second part of the section. Some 
details about Green's functions and the derivation of the partition function are relegated
to the Appendix.

To summarise, we compute the superstring partition function on the annulus in the presence of background
open string tachyonic fields 
and show in particular that the \RR sector contributes non-trivially. In the process we 
construct
boundary states in the presence of linear tachyons; as a corrolary we compute the WZ
couplings of non-BPS D-branes. Furthermore we discuss the quantisation of an open string in the 
presence of
a background tachyon, comment on the fate of the open string GSO projection 
and open/closed string duality.

\section{The superstring action in a tachyon and gauge field background}\label{sec1a}
\setcounter{equation}{0}

The world sheet action for the superstring in the background of a tachyon and abelian gauge field is
\begin{equation}
S=S_{\mbox{\scriptsize bulk}}+S_{\mbox{\scriptsize bndy}},
\end{equation}
with the standard NSR action in the bulk (we set $\a'=2$)
\begin{equation}
S_{\mbox{\scriptsize bulk}}=\frac{1}{4\pi}\int_{\S}\,\rmd^2z\left(\p_zX^{\m}\p_{\bz}X_{\m}+ 
\psi^{\m}\p_{\bz}\psi_{\m}+\tilde{\psi}^{\m}\p_z\tilde{\psi}_{\m}\right),
\end{equation}
and the boundary action in superspace~\cite{KMM2},~\cite{KL}
\begin{equation}
S_{\mbox{\scriptsize bndy}}=-\frac{1}{2\pi}\int_{\p\S}\rmd s\rmd\Theta\,\left(\G D\G+T({\bf X})\G+
\frac{i}{2}A_{\m}({\bf X})D{\bf X}^{\m}
\right).
\end{equation}
The boundary superspace coordinates are $(s,\Theta)$, 
where $\Theta$ is the boundary Grassmann 
coordinate and $s=r\phi$, $\phi$ being the angular coordinate on the boundary. 
Here ${\bf X}=X+\Theta\theta$, where $\theta$ is a boundary fermion, whose precise 
relation to the bulk fermions $\psi$ and 
$\tilde{\psi}$ will be determined below and 
$D=\p_{\Theta}+\Theta\p_s$. $\G=\r+\Theta K$ is an auxiliary 
boundary superfield \cite{WittenK,HKM}. 

The world-sheet $\S$ is the annulus with inner radius 
$a<1$ and outer radius equal to one. In 
this case there is generically an independent set of background and auxiliary fields 
on each component of the boundary, 
though to avoid cluttering of notation this is not explicitly indicated in the above. 
Performing the integral over $\Theta$ and 
integrating out the auxiliary field $K$ one obtains
\begin{equation}
S_{\mbox{\scriptsize bndy}}=
-\frac{1}{2\pi}\int_{\p\S}\rmd s\left(\dot{\r}\r+\p_{\m}T(X)\theta^{\m}\r-\frac{1}{4}T(X)^2
+\frac{i}{2}A_{\m}(X)\dot{X}^{\m}+\frac{i}{4}F_{\m\n}(X)\theta^{\m}\theta^{\n}\right),
\end{equation}
where $\dot{\r}=\p_s\r$ {\em etc.} The world-sheet theory is exactly solvable 
in the presence of a linear tachyon profile and a constant abelian field strength
\begin{equation}
T(X)=u_{\m}X^{\m},\qquad A_{\m}(X)=-\frac{1}{2}F_{\m\n}X^{\n}.
\end{equation}
In this case the boundary action is
\begin{equation}\label{bndyaction}
S_{\mbox{\scriptsize bndy}}
=\frac{1}{8\pi}\int_{\p\S}\rmd s\left(u_{\m\n}X^{\m}X^{\n}+iF_{\m\n}\p_sX^{\m}X^{\n}
-iF_{\m\n}\theta^{\m}\theta^{\n}-4\p_s\r\r-4u_{\m}\theta^{\m}\r\right),
\end{equation}
where we defined $u_{\m\n}\equiv u_{\m}u_{\n}$.

The relationship between the boundary and bulk fermions can be determined as follows. 
On-shell the variation of the 
fermionic bulk term reduces to the boundary contribution 
\begin{equation}
\d S_{\mbox{\scriptsize bulk}}=-\frac{i}{4\pi}\int\,\rmd s
\left(\psi^{\m}(s)\d\psi_{\m}(s)+\tilde{\psi}^{\m}(s)\d\tilde{\psi}_{\m}(s)\right),
\end{equation}
where we took into account the transformation of the bulk fermions from $z=re^{i\phi}$ to $s=r\phi$ 
variables. As above we introduce the boundary action with the boundary fermion $\theta$, 
which is a {\em new} field
and relate it to the bulk fermions by treating it as the Lagrange multiplier 
in the following modified boundary action
\begin{equation}
S'_{\mbox{\scriptsize bndy}}=S_{\mbox{\scriptsize bndy}}-\frac{i}{8\pi}\int_{\p\S}\rmd s\,\theta^{\m}
\bigl(\psi_{\m}-i\eta\tilde{\psi}_{\m}\bigr)\,,
\end{equation}
with $\eta=\pm 1$.
As we will see in a moment this (and a similar choice $\tilde{\eta}=\pm 1$ 
on the other component of the 
boundary) corresponds to the spin structure,
 since it leads to different ways of 
identifying the boundary fermion in 
terms of the bulk fermions.\footnote{Note that the equation of motion for $\theta$ 
reduces to the standard relation 
$\psi=i\eta\tilde{\psi}$ in the case of vanishing background fields.}
Introducing $\psi_{\pm}=\psi\pm i\eta\tilde{\psi}$ the variation coming from the 
fermionic parts of the action now reads (on the $r=1$ boundary)
\begin{equation}
\d S =-\frac{i}{8\pi}\int\,\rmd\phi\Bigl[
\psi_{-}\d\psi_{+}+\bigl(\psi_{+}+\theta\bigr)\d\psi_{-}
-\psi_{-}\d\theta\Bigr]+\d S_{\mbox{\scriptsize bndy}}\,.
\end{equation}
We define the boundary conditions for fermions by requiring the variation of the total 
action to vanish on-shell.  
Since $\d\psi_{-}$ and $\d\theta$ are independent variables on the boundary
the vanishing of the coefficient of $\d\psi_{-}$ yields
\begin{equation}\label{theta}
-\theta=\psi_{+}=\psi+i\eta\tilde{\psi},
\end{equation}
\ie\ it now relates the bulk and boundary fermions. This relation 
implies $\d\psi_+=-\d\theta$ and the remaining part of $\d S$ gives 
the boundary conditions for $\theta$ (\cf Section~\ref{sec2}).

The choice of boundary fermion $-\theta=\psi+i\eta \tilde{\psi}$ 
in terms of bulk fermions is precisely the 
choice of spin structure. Since 
we have two boundaries for the annulus we have in sum four different 
possibilities to identify the boundary fermions 
with the bulk fermions (\cf~\cite{PC}). These cases should be combined 
with the conditions for bulk fermions to be 
antiperiodic or periodic around the circle, so together we would get 
eight different sectors. However, as we will show in Section~\ref{sec2},
there are only four different sectors since the spin structure enters in the 
final answers only through the combination 
$\eta\tilde{\eta}$.

\section{The annulus partition function via Green's functions}\label{sec2}
\setcounter{equation}{0}

In this section we determine the complete partition function in the closed 
channel by the method of Green's functions~\cite{abou}. 
For the sake of clarity we will analyse the different sectors separately and summarise the 
results here. The computational details are presented in the Appendix.

\subsection{The bosonic sector}

Besides the Laplace equation in the bulk the $X^\mu$'s satisfy\footnote{The 
normal and tangential derivatives are
$\p_r=\mbox{\scriptsize $\frac{1}{|z|}$}(z\p_z+\bz\p_{\bz})\,,$ 
$\p_s=\mbox{\scriptsize $\frac{i}{|z|}$}(z\p_z-\bz\p_{\bz})\,,$ respectively.}
\ba\label{bc}
(z\p_z+\bz\p_{\bz})X_{\m}+u_{\m\n}X^{\n}+F_{\m\n}(z\p_z-\bz\p_{\bz})X^{\n}&=&0,\qquad |z|=1\,,
\nonumber \\
-(z\p_z+\bz\p_{\bz})X_{\m}+av_{\m\n}X^{\n}+L_{\m\n}(z\p_z-\bz\p_{\bz})X^{\n}&=&0,\qquad |z|=a\,,
\ea
where $z=re^{i\phi}$, $L_{\mu\nu}$ and $v_{\mu\nu}=v_\mu v_\nu$ are respectively 
the gauge field and tachyon on the $r=a$  boundary. 
The Green's function corresponding to these boundary conditions is 
\ba
\label{Gb}
G(z,w)&=&-\ln|z-w|^2+A-\frac{1}{2}Cu\ln|z|^2\ln|w|^2+C\ln|z|^2+C^T\ln|w|^2\nonumber\\
& &+\,\,\sum_{k>0}\bigl(\a_{k}z^k+\a_{-k}z^{-k}+\tilde{\a}_{k}\bz^k+\tilde{\a}_{-k}\bz^{-k}\bigr)\,,
\ea
where  
\ba
A&=&2(1-av\ln a)(u+av-auv\ln a)^{-1},\\
C&=&av(u+av-auv\ln a)^{-1}.
\ea
The explicit expressions for the oscillators are given in the Appendix . $G(z,w)$ satisfies 
$G_{\m\n}(z,w)=G_{\n\m}(w,z)$ and is in fact the propagator 
$\la X_{\m}(z)X_{\n}(w)\ra$. 

One finds the bosonic contribution to the partition function 
(up to normalization) by differentiating the boundary action~(\ref{bndyaction})
with respect to the world-sheet 
couplings~\cite{WittenBSFT} and using this Green's function (for more details see the Appendix). 
The resulting expression is 
\ba
\label{Zb}
Z_{\mbox{\scriptsize bos}}(a)&=&\det(u+av-auv\ln a)^{-1/2}\prod_{k=1}^{\infty}\det(1+u/k+F)^{-1}\det(1+av/k+L)^{-1}
\nonumber\\&&\times\,\,
\det\left(1-a^{2k}S_k(u,F)S_k(av,L)\right)^{-1}\,,
\ea
where 
\begin{equation}
S_k(u,F)=\frac{k-u-kF}{k+u+kF}\,.
\end{equation}
This result is purely formal and has to be regularized.
The infinite product above diverges and is treated in Section~\ref{b2} 
(\cf equations~(\ref{nsnsnorm}) and~(\ref{rrnorm}) below). As in the disc case the bosonic
and fermionic divergences combine to produce a finite result~\cite{KMM2,Andreev,KL}. Furthermore
the matrix $(u+av-auv\ln a)$ has rank one or two depending on whether the tachyons $u_\mu\,,v_\mu$ 
are switched on in one or more directions.\footnote{One can always rotate to reduce $u$ and $v$ to 
two-dimensional vectors $(u_1,0)$ and $(v_1,v_2)$.} The determinant above should be
understood as a product of the determinant of the maximal rank sub-matrix and the regularised volume
of the remaining space-time directions. 

In principle there may also be an overall dependence on the modulus 
$a$ in the partition function that could not be fixed by the previous considerations. 
By looking at the change of the 
action under variations of the modulus~\cite{abou} one can also derive an equation for $\p_a\ln Z$. 
It turns out that this equation is consistent with the above 
expression~(\ref{Zb}) for the partition function, \ie\ no extra dependence on $a$ 
appears (for details see the Appendix). Note also that in the 
limit $a\to 0$ one recovers the (bosonic) 
partition function on the disc (setting $L=0$)~\cite{WittenBSFT}. 

The partition function obtained above agrees with the one computed in section~\ref{sec3} 
using the boundary state formalism. When comparing the two results one should note that the correct
integration measure on the annulus is $\frac{da}{a^2}$.

\subsection{The \NSNS sector}

In the \NSNS sector neither $\r$ nor $\theta$ have zero-modes and the auxiliary boundary 
fermion $\r$ can be 
integrated out. The fermionic part of the boundary action becomes 
\begin{equation}
{S'}^F_{\mbox{\scriptsize bndy}}=\frac{1}{8\pi}\int_{\p\S}\left(u_{\m\n}
\theta^{\m}\p_s^{-1}\theta^{\n}-iF_{\m\n}\theta^{\m}\theta^{\n}
-i\theta^{\m}\bigl(\psi_{\m}-i\eta\tilde{\psi}_{\m}\bigr)\right),
\end{equation}
and the fermionic boundary conditions following from the (on-shell) vanishing 
of the total variation of $S_{\mbox{\scriptsize bulk}}+
S'_{\mbox{\scriptsize bndy}}$ (\cf the discussion in Section~\ref{sec1a}) are 
\ba\label{bcf}
\left(\d_{\m\n}+iu_{\m\n}\p_s^{-1}+F_{\m\n}\right)\psi^{\n}&=&i\eta
\left(\d_{\m\n}-iu_{\m\n}\p_s^{-1}-F_{\m\n}\right)\tilde{\psi}^{\n},\qquad r=1,\nonumber \\ 
\left(\d_{\m\n}-iav_{\m\n}\p_s^{-1}-L_{\m\n}\right)\psi^{\n}&=&i\tilde{\eta}
\left(\d_{\m\n}+iav_{\m\n}\p_s^{-1}+L_{\m\n}\right)\tilde{\psi}^{\n},\qquad r=a\,.
\label{bdrcondgreen}
\ea
Introducing the four kinds of Green's functions on the boundaries
\ba
G_{++}(z,w)&\equiv&\la\psi(z)\psi(w)\ra=-i\frac{\sqrt{zw}}{z-w}+
\sum_{r=1/2}^{\infty}\bigl(\psi_r(w)z^r+\psi_{-r}(w)z^{-r}\bigr)\,,\nonumber \\
G_{--}(\bz,\bw)&\equiv&\la\tilde{\psi}(\bz)\tilde{\psi}(\bw)\ra=-i\frac{\sqrt{\bz\bw}}{\bz-\bw}+
\sum_{r=1/2}^{\infty}\bigl(\tilde{\psi}_r(\bw)\bz^r+\tilde{\psi}_{-r}(\bw)z^{-r}\bigr)\,,
\nonumber \\
G_{+-}(z,\bw)&\equiv&\la\psi(z)\tilde{\psi}(\bw)\ra=
\sum_{r=1/2}^{\infty}\bigl(a_r(\bw)z^r+a_{-r}(\bw)z^{-r}\bigr)\,,\nonumber\\
G_{-+}(\bz,w)&\equiv&\la\tilde{\psi}(\bz)\psi(w)\ra=
\sum_{r=1/2}^{\infty}\bigl(b_r(w)\bz^r+b_{-r}(w)\bz^{-r}\bigr)\,,
\ea
the boundary conditions on the Green's functions can be written as
\ba
\left(1+F)z\p_z+u\right)G_{+\pm}+i\eta\left((1-F)\bz\p_{\bz}+u\right)
G_{-\pm}&=&0,\qquad |z|=1,\nonumber \\
\left(-(1-L)z\p_z+av\right)G_{+\pm}+i\tilde{\eta}\left(-(1+L)\bz\p_{\bz}+av\right)
G_{-\pm}&=&0,\qquad |z|=a.
\ea
The boundary fermion at $r=1$ is related to the bulk 
fermions by $\theta=-(\psi+i\eta\tilde{\psi})$ (\cf equation~(\ref{theta})) and therefore the 
propagator is 
\begin{equation}
\la\theta\theta\ra=G_{++}-G_{--}+i\eta\bigl(G_{+-}+G_{-+}\bigr)\,.
\end{equation}
On the boundary $r=a$ the second boundary fermion $\tilde{\theta}$ is
\begin{equation}
\tilde{\theta}=-(\psi+i\tilde{\eta}\tilde{\psi})
\end{equation}
and consequently has propagator
\begin{equation}
\la\tilde{\theta}\tilde{\theta}\ra=G_{++}-G_{--}+i\tilde{\eta}\bigl(G_{+-}+G_{-+}\bigr)\,.
\end{equation}
A straightforward, though tedious calculation determines the oscillators 
of the Green's functions (whose explicit
expressions are again collected in the Appendix) and, using the expressions for 
the boundary fermion propagators, one finds that 
the resulting contribution to the partition function from the NS-NS sector spin structures is 
formally 
\begin{equation}\label{ZfNS}
Z_{\mbox{\scriptsize \NSNS}}(a,\eta\tilde{\eta})=\prod_{r=1/2}^{\infty}\det(1+u/r+F)
\det(1+av/r+L)
\det\left(1-\eta\tilde{\eta}a^{2r}S_r(u,F)S_r(av,L)\right)\,.
\end{equation}
Due to the closed string GSO projection the contributions from different spin structures 
($\eta{\tilde\eta}=\pm 1$) should be added with the opposite sign. This removes the 
closed string tachyon (\cf Section~\ref{sec3}). 

\subsection{The \RR sector}\label{sec33}

The \RR sector is more subtle, due to the appearance of the 
$\r$ and $\theta$ zero-modes. Since the zero-mode drops out of the kinetic term of the 
auxiliary boundary fermion $\r$ one cannot integrate 
out $\r$ completely as in the \NSNS sector. Instead we will integrate out the non-zero 
modes and treat the zero-modes separately. Then the boundary 
condition on the non-zero modes of $\psi$, ${\tilde\psi}$ are exactly as in~(\ref{bcf}). 
The Green's functions now read  
\ba
G_{++}(z,w)&\equiv&\la\psi(z)\psi(w)\ra\nonumber \\
&=&-i\frac{1}{z-w}\left(w\Theta(|z|-|w|)+z\Theta(|w|-|z|)\right)
+\sum_{r=1}^{\infty}\bigl(\psi_r(w)z^r+\psi_{-r}(w)z^{-r}\bigr)\,,\nonumber \\
G_{--}(\bz,\bw)&\equiv&\la\tilde{\psi}(\bz)\tilde{\psi}(\bw)\ra\nonumber \\
&=&-i\frac{1}{\bz-\bw}
\left(\bw\Theta(|z|-|w|)+
\bz\Theta(|w|-|z|)\right)+\sum_{r=1}^{\infty}\bigl(\tilde{\psi}_r(\bw)\bz^r+
\tilde{\psi}_{-r}(\bw)z^{-r}\bigr)\,,\nonumber \\
G_{+-}(z,\bw)&\equiv&\la\psi(z)\tilde{\psi}(\bw)\ra=\sum_{r=1}^{\infty}\bigl(a_r(\bw)z^r+
a_{-r}(\bw)z^{-r}\bigr)\,,\nonumber \\
G_{-+}(\bz,w)&\equiv&\la\tilde{\psi}(\bz)\psi(w)\ra=\sum_{r=1}^{\infty}\bigl(b_r(w)\bz^r+
b_{-r}(w)\bz^{-r}\bigr)\,,
\ea
where $\Theta(|z|-|w|)$ is the step function.
A completely analogous calculation to the one for the \NSNS sector shows that the contribution 
of the \RR non-zero modes 
to the partition function is 
\begin{equation}\label{ZfRRnz}
Z_{\mbox{\scriptsize \RR}}(a,\eta\tilde{\eta})=\prod_{r=1}^{\infty}\det(1+u/r+F)\det(1+av/r+L)\det
\left(1-\eta\tilde{\eta}a^{2r}S_r(u,F)S_r(av,L)\right).
\end{equation}
For the zero-modes the kinetic terms of the auxiliary boundary fermions $\r$, 
$\tilde{\r}$ are absent and the relevant part of the boundary action reads\footnote{For 
the \RR zero-modes we impose the free boundary conditions, 
thereby relating the left and right movers.}
\begin{equation}
S_{\mbox{\scriptsize bndy}}^{(0)}=-\frac{1}{8\pi}\int_{\p\S}ds\,\bigl(4u_{\m}\theta_0^{\m}\r_0+
iF_{\m\n}\theta_0^{\m}\theta_0^{\n}\bigr)\,.\label{actfermzero}
\end{equation}
Integrating out $\r_0$ and $\tilde{\r}_0$ we have 
\begin{equation}
Z_{\mbox{\scriptsize \RR}}^{(0)}=\left\la\left(u_{\m}\theta_0^{\m}
\exp(\frac{i}{4}F_{\r\l}\theta^{\r}_0\theta^{\l}_0)\right)_{r=1}
\left(av_{\m}\tilde{\theta}_0^{\m}\exp(\frac{i}{4}L_{\r\l}
\tilde{\theta}^{\r}_0\tilde{\theta}^{\l}_0)\right)_{r=a}
\right\ra.
\end{equation}
Translating this into Hilbert space language, we see that the zero-mode 
part of the partition function in the \RR sector 
is given by the amplitude of the ``in-states'' at $r=a$ and the ``out-states'' at $r=1$. 
The explicit expression for the in- and out-states will be derived in the following. 
The zero-modes of the bulk fermions satisfy
\begin{equation}
\{\psi_0^{\m},\psi_0^{\n}\}=\d^{\m\n}=\{\tilde{\psi}_0^{\m},\tilde{\psi}_0^{\n}\},\qquad
\{\psi_0^{\m},\tilde{\psi}_0^{\n}\}=0.
\end{equation}
The action of  $\psi_0^{\m}$ and $\tilde{\psi}_0^{\m}$ on the \RR vacuum (in a non-chiral basis)
\begin{equation}
|A,\tilde{B}\ra\equiv\lim_{z,\bz\to0}S^A(z)\tilde{S}^B(\bz)|0\ra,\quad A,B=1,\ldots,32
\end{equation}
can be realized as\footnote{We follow here the approach of~\cite{Frauferm}.}
\ba
\psi_0^{\m}|A,\tilde{B}\ra&=&\frac{1}{\sqrt{2}}{(\G^{\m})^A}_C{(1)^B}_D|C,\tilde{D}\ra\,,\nonumber \\
\tilde{\psi}_0^{\m}|A,\tilde{B}\ra&=&\frac{1}{\sqrt{2}}{(\G_{11})^A}_C{(\G^{\m})^B}_D|C,\tilde{D}\ra\,.
\ea
The vacuum `in-state' is defined by the free boundary condition
\begin{equation}
\bigl(\psi_0^{\m}-i\tilde{\eta}\tilde{\psi}_0^{\m}\bigr)|-\tilde{\eta}\ra=0.
\end{equation}
Explicitly it is 
\begin{equation}
|-\tilde{\eta}\ra={\cal M}^{(\tilde{\eta})}_{AB}|A,\tilde{B}\ra,\qquad {\cal M}^{(\tilde{\eta})}_{AB}=
\left[C\G_{11}\frac{1+i\tilde{\eta}\G_{11}}{1+i\tilde{\eta}}\right]_{AB},
\end{equation}
where  $C$ is the charge conjugation matrix. Similarly, the vacuum ``out-state'' is 
\begin{equation}
\la\eta|=\la A,\tilde{B}|{\cal N}^{(\eta)}_{AB},\qquad 
{\cal N}^{(\eta)}_{AB}=-\left[C\G_{11}\frac{1-i\eta\G_{11}}{1+i\eta}\right]_{AB},
\end{equation}
and satisfies
\begin{equation}
\la\eta|\bigl(\psi_0^{\m}+i\eta\tilde{\psi}_0^{\m}\bigr)=0\,.
\end{equation}
We have
\begin{equation}
\la\eta|-\tilde{\eta}\ra=-32\d_{\eta,-\tilde{\eta}}\,,
\end{equation}
so that, as usual, only one of the two spin structure contributions of the \RR sector is non-zero. 
Since the boundary fermions anti-commute the expansion of the exponentials will in general 
terminate at fourth order in the gauge field strengths and we have
\begin{equation}
Z_{\mbox{\scriptsize\RR}}^{(0)}=au_{\m}v_{\n}\la\theta^{\m}_0|\tilde{\theta}_0^{\n}\ra-\frac{1}{16}
au_{\m}v_{\n}F_{\r\l}L_{\s\t}
\la\theta^{\m}_0\theta^{\r}_0\theta^{\l}_0|\tilde{\theta}_0^{\n}\tilde{\theta}_0^{\s}
\tilde{\theta}_0^{\t}\ra+\cdots
\end{equation}
Since $\tilde{\theta}_0^{\m}$ and ${\theta}_0^{\m}$ act as creation and annihilation 
operators on the in and out vacua 
respectively, only terms with the same number of $\theta_0^{\m}$ and $\tilde{\theta}_0^{\m}$ 
give a non-zero 
contribution. We explicitly compute this in two particular cases. First consider turning on 
tachyons $u\,,v$ in directions transversal to a gauge field $L=F$. In this case the zero-modes
contribute as
\be
Z_{\mbox{\scriptsize\RR}}^{(0)}\sim au_{\m}v^{\m}\det(1+F)\,.
\ee
Next consider again $L=F$ restricted to, say, four directions $\mu=1,2,3,4$ with
the tachyons non-zero in the same directions. As long
as the tachyons $u,v$ are general we may in fact rotate $F$ to bring it to a block-diagonal 
form consisting of two antisymmetric 
$2\times 2$ matrices with independent entries $f_1,f_2$. After some algebra the result becomes
\be
Z_{\mbox{\scriptsize\RR}}^{(0)}\sim au_iv^i+af_1^2(u^3v^3+u^4v^4)+af_2^2(u^1v^1+u^2v^2)\,.
\ee
Covariantly this is written as 
\be
Z_{\mbox{\scriptsize\RR}}^{(0)}\sim 
au_{\m}v^{\m}+au^{\m}F^2_{\m\n}v^{\n}-\frac{a}{2}u_{\m}v^{\m}F_{\r\l}F^{\r\l}
\ee
and indicates a mixing of the tachyons with the gauge field in the space-time effective action. 

Summarising, the contribution of the \RR sector to the full partition function in 
the closed string channel is
\ba
Z_{\mbox{\scriptsize\RR}}(a)=Z_{\mbox{\scriptsize\RR}}^{(0)}
\prod_{r=1}^{\infty}\det(1+u/r+F)\det(1+av/r+L)
\det\left(1-\eta\tilde{\eta}a^{2r}S_r(u,F)S_r(av,L)\right)\,,\nonumber \\
\ea
where due to the zero-modes only the spin structure $\eta\tilde{\eta}=-1$ gives a non-vanishing 
contribution. The above expression should contribute with an overall minus sign
to the total partition function
so as to respect open/closed string duality. This is discussed in more detail in Section~\ref{sec3}.

\section{Boundary states in the presence of a tachyon}\label{sec3}
\setcounter{equation}{0}

In this section we construct coherent states which represent D-branes in 
the presence of a boundary tachyon perturbation. These states are first obtained as
solutions of the boundary conditions viewed as an eigenvector equation; the functional
approach~\cite{CLNY} is then used to normalise them. Finally the cylinder diagram is computed for a 
closed string propagating between two parallel D-branes with tachyon perturbations turned on.

\subsection{Boundary states as eigenvector solutions of boundary conditions}\label{b1}

In this sub-section we construct the boundary state for an unstable D9-brane in 
Type IIA in the presence of a linear tachyon. 
Our solution will be complete apart from normalisation, since the 
boundary state shall be obtained, in the usual way, by interpreting the boundary conditions 
as eigenvector equations. We construct a boundary state as an eigenvector satisfying at $\htau=0$
\ba\label{bccyl}
\p_{\htau}X^9+\hu X^9&=&0\,, \nonumber \\
\p_{\htau}\psi^9+\hu\psi^9&=& i\eta(\partial_\tau{\tilde\psi}^9-\hu{\tilde\psi}^9)\,,
\ea
where $\eta$ corresponds to the two spin structures each in the \NSNS and \RR sectors and $\hu$ is 
some constant. Comparing with boundary conditions~(\ref{bdrbc}) we obtain the boundary 
states relevant to us by taking $\hu=u^2,-e^lv^2$ respectively. This identification is consistent 
with the boundary conditions~(\ref{bdrcondgreen}) on the annulus with no gauge fields and tachyon only in 
one direction. The boundary conditions~(\ref{bccyl}) in modes read
\be
2ip^9+\hu x^9=0\,,\qquad 
(n+\hu)\alpha^9_n=(n-\hu){\tilde\alpha}^9_{-n} \,,\qquad
(r+\hu)\psi^9_r=i\eta(r-\hu){\tilde\psi}^9_{-r} \,,
\ee
for $n=\pm 1\,,\pm 2\,,\dots$, $r=\pm\half\,,\pm\frac{3}{2}\,,\dots$ in the \NSNS sector and  
$r=\pm 1\,,\pm 2\,,\dots$ in the 
\RR sector. We shall discuss the bosonic and \RR sector zero-modes below. 
The coherent state which solves these equations is
\ba
\ket{B,\hu,\eta}_{\mbox{\scriptsize\NSNS,\RR}}&=&{\cal N}_{\mbox{\scriptsize\NSNS,\RR}}(\hu)
\exp\left(\sum_{n=1}^\infty\frac{1}{n}\frac{n-\hu}{n+\hu}
\alpha^9_{-n}{\tilde\alpha}^9_{-n}+i\eta\sum_{r>0}^\infty\frac{r-\hu}{r+\hu}
\psi^9_{-r}{\tilde\psi}^9_{-r}\right)\nonumber \\ &&\times\,\,
\ket{B,\hu,\eta}_{\mbox{\scriptsize\NSNS,\RR}}^{(0)}\ket{B\mbox{ other}}_{\mbox{\scriptsize\NSNS,\RR}}\,.
\ea
Here ${\cal N}_{\mbox{\scriptsize\NSNS,\RR}}(\hu)$ is the $\hu$ dependent normalisation, 
$\ket{B,\hu,\eta}^{(0)}$ contains the zero-mode dependence (see below) and 
$\ket{B\mbox{ other}}$ is the contribution of the other (Neumann) directions. 
With the present boundary conditions the images of fields outside the disc 
are rather complicated; for example for the world-sheet fermion
\be
\psi^9(z)=i\eta\frac{\partial_\tau+\hu}{\partial_\tau-\hu}{\tilde\psi}^9(1/{\bar z})\,.\label{image}
\ee
The bosonic zero-mode conditions
\be
(2ip+\hu x)\ket{B,\hu}^{(0)}_{\mbox{\scriptsize bosonic}}=0\,,
\ee
are solved by
\be
\ket{B,\hu}^{(0)}_{\mbox{\scriptsize bosonic}}=
e^{-\frac{1}{4}\hu x^2}\ket{0}_p\,,\label{boszero}
\ee
in the momentum basis. To treat the \RR sector fermionic zero modes 
we define in the usual fashion (see for example~\cite{S3})
\be
\psi^9_\eta=\frac{1}{\sqrt{2}}(\psi^9_0+i\eta{\tilde\psi}^9_0)\,,
\ee
which satisfy for non-zero $\hu$ the (Dirichlet) boundary condition (see equation~(\ref{image}))
\be
\psi^9_\eta\ket{B,\hu,-\eta}_{\mbox{\scriptsize\RR}}=0\,,\qquad\hu\neq 0\,.
\label{ferzero}
\ee

In the \NSNS sector there are no fermionic zero-modes
and requiring closed string GSO invariance produces a unique boundary state 
\be
\ket{B,\hu}_{\mbox{\scriptsize\NSNS}}=\half(\ket{B,\hu,+}_{\mbox{\scriptsize\NSNS}}-
\ket{B,\hu,-}_{\mbox{\scriptsize\NSNS}})\,.
\ee
Similarly, in the \RR sector the GSO invariant boundary state is 
\be
\ket{B,\hu}_{\mbox{\scriptsize\RR}}=2i(\ket{B,\hu,+}_{\mbox{\scriptsize\RR}}+
\ket{B,\hu,-}_{\mbox{\scriptsize\RR}})\,,\qquad\hu\neq 0\,.
\ee

\subsection{Normalisation of the boundary states}\label{b2}

In principle the normalisation of the above constructed boundary states
can be fixed by computing the cylinder amplitude and 
performing a modular transformation to compare with the one
loop open string partition function. As we will see it is quite difficult to 
determine the modular properties of the functions obtained in the cylinder 
channel directly. A second way~\cite{CLNY} involves integrating out the
boundary degrees of freedom. This approach has been used in~\cite{da} to normalise 
the \NSNS boundary state in the presence of a tachyon. We review briefly the considerations
of~\cite{CLNY} and apply them to the problem at hand.  Firstly one must
set up a complete orthonormal set of bosonic and fermionic coordinates. Define
\be
{\bar x}^\mu_m = a^{\mu\dagger}_m+{\tilde a}^\mu_m\,,\qquad
x^\mu_m = a^\mu_m+{\tilde a}^{\mu\dagger}_m \,,
\ee
with $m>0$. Together with $q^\mu$, the centre of mass position,
this gives a complete commuting set of bosonic coordinates. The state 
\be
\ket{x,{\bar x}}=\exp\left\{-\frac{1}{2}({\bar x}|x)-(a^\dagger|{\tilde 
a}^\dagger) +(a^\dagger|x)+({\bar x}|{\tilde a}^\dagger)\right\}\ket{0}\,,\label{373}
\ee
satisfies the eigenvector equation
\ba
\left[a^{\mu\dagger}_m+{\tilde a}^\mu_m -{\bar x}^\mu_m\right]
\ket{x,{\bar x}}&=&0\,,\\
\left[a^\mu_m+{\tilde a}^{\mu\dagger}_m - x^\mu_m\right]
\ket{x,{\bar x}}&=&0\,.
\ea
In the above
\be
({\bar x}|x) = \sum_{\mu=1}^{10}\sum_{m=1}^\infty {\bar x}^\mu_m x_{m,\mu}\,.\label{374}
\ee
The states $\ket{x,{\bar x}}$ are complete as can be seen from
\be
\int\!{\cal D}x{\cal D}{\bar x}\ket{x,{\bar x}}\bra{x,{\bar x}}=1\,.
\ee
For fermions one defines
\be
\psi^\mu+i\eta\tilde{\psi}^\mu\equiv\theta^\mu\equiv\sum_n\theta^\mu_n 
e^{-in\sigma}
\ee
and
\be
{\bar\theta}^\mu_n\equiv\theta^{\mu\dagger}_{-n}\,.
\ee
These anti-commute in the usual fashion
\be
\{\theta^\mu_m,\theta^\nu_n\}=0\,.
\ee
Ignoring for the time being the \RR zero-modes we look for
eigenvectors satisfying
\ba
({\bar\theta}^\mu_m-\psi^{\mu\dagger}_m- i\eta{\tilde\psi}^\mu_m)\ket{\theta,
{\bar\theta};\eta}&=&0\,, \\
(\theta^\mu_m-\psi^\mu_m+i\eta{\tilde\psi}^{\mu\dagger}_m)\ket{\theta,
{\bar\theta};\eta}&=&0 \,.
\ea
They are
\be
\ket{\theta,{\bar\theta};\eta}=
\exp\left\{-\frac{1}{2}({\bar\theta}|\theta)+ 
i\eta(\psi^\dagger|{\tilde\psi}^\dagger)+({\tilde\psi}^\dagger|\theta)-i\eta 
({\bar\theta}|{\tilde\psi}^\dagger)\right\}\ket{0;\eta}\,,\label{385}
\ee
and satisfy the completeness relations
\be
\int{\cal D}{\bar\theta}{\cal D}\theta
\ket{\theta,{\bar\theta};\eta}\bra{\theta,{\bar\theta};\eta}=1\,.
\ee
The inclusion of bosonic and fermionic zero-modes are discussed in detail in~\cite{CLNY} 
which should be consulted by the interested reader. Here we simply point out that these
act directly on the zero field vacuum and are not integrated over. The boundary state 
can be written as
\be
\ket{B,\hu,\eta}=\int{\cal D}{\bar x}{\cal D}x{\cal D}{\bar\theta}{\cal D}\theta
e^{-S(x,{\bar x},q;\theta,{\bar\theta},\theta_0)}\ket{x,{\bar x}}\ket{\theta,{\bar\theta},\eta}\,.
\ee
where $S$ in our case is the boundary action for the linear tachyon. Evaluating the functional
integrals explicitly yields
\ba
\ket{B,\hu,\eta}_{\mbox{\scriptsize\NSNS, \RR}}&=&{\cal N}_{\mbox{\scriptsize\NSNS,\RR}}
\frac{\prod_{r>0}^\infty(1+\frac{\hu}{r})}{\prod_{n=1}^\infty(1+\frac{\hu}{n})}
\exp\left(\sum_{n=1}^\infty\frac{1}{n}\alpha^\mu_{-n}S^n_{\mu\nu}{\tilde\alpha}^\mu_{-n}\right)
\nonumber \\
&&\times\exp\left(i\eta\sum_{r>0}^\infty\psi^\mu_{-r}S^r_{\mu\nu}{\tilde\psi}^\mu_{-r}\right)
\ket{B,\hu,\eta}_{\mbox{\scriptsize\NSNS, \RR}}^{(0)}
\ket{B\mbox{ other}}_{\mbox{\scriptsize\NSNS,\RR}}
\,, \nonumber \\
\ea
where $r$ is half-integral in the \NSNS sector and integral in the \RR sector, 
and the zero-mode part of the boundary state is as discussed above. ${\cal N}$ is the  
normalisation of the zero-mode part of the boundary state (and the other directions). 
The matrix $S$ is 
\be
S^n_{\mu\nu}=\mbox{diag}(-1,\dots,-1,1,\dots,1,\frac{n-\hu}{n+\hu})\,,
\ee
with entries $-1,1$ in the Neumann, Dirichlet directions, 
respectively. In the \RR sector the above infinite products cancel  
between the bosons and fermions, while in the \NSNS sector they need to be 
regularised. This gives the $\hu$ dependence of the normalisation as~\cite{KMM2}
\be
{\cal N}_{\mbox{\scriptsize\NSNS}}(\hu)=\frac{1}{2}\hu 4^{\hu}B(\hu,\hu)
{\cal N}_{\mbox{\scriptsize\NSNS}}
\ee
with $B$ the Euler Beta function and ${\cal N}_{\mbox{\scriptsize\NSNS}}$ is a $\hu$ independent 
constant.\footnote{A similar infinite product was also encountered in Section~\ref{sec2} 
and should be treated in an analogous fashion.}

Viewed as an eigenvector the boundary state thus obtained agrees with the one constructed 
in sub-section~\ref{b1}. Further, we have determined the normalisation of the non-zero mode part by 
integrating out the boundary degrees of freedom. The bosonic and \RR fermionic zero-modes are not
integrated; instead they act directly on the closed string vacuum~\cite{CLNY}. In particular 
the bosonic zero-mode's action is
\be
\exp\left(-\frac{1}{4}\hu x^2\right)\ket{0}_p\,,
\ee
fixing the normalisation of equation~(\ref{boszero}). The action of the fermionic zero-mode 
is discussed in the paragraph around equation~(\ref{actfermzero}) giving the zero-mode part of
the D9-boundary state in the presence of a tachyon as
\be
\sqrt{\hu}\psi^9_{\eta}\ket{B9,-\eta}_{\mbox{\scriptsize\RR}}^{(0)}\,,\label{zeroactrr}
\ee
where $\ket{B9,\eta}_{\mbox{\scriptsize\RR}}^{(0)}$ is the zero-mode part of the usual D9-brane
\RR boundary state (\cf~\cite{S3}). This fixes the normalisation of equation~(\ref{ferzero}).
The normalisation constant ${\cal N}_{\mbox{\scriptsize\NSNS}}$ 
is\footnote{With this normalisation the \NSNS boundary state reduces in the two limits 
$\hu\to0$, $\hu\to\infty$ to the  usual \NSNS boundary states 
for a non-BPS D9-brane and a BPS D8-brane of Type IIA with no background tachyon, respectively.} 
\be
{\cal N}_{\mbox{\scriptsize\NSNS}}=T_{\mbox{\scriptsize non-BPS D9}}\,.\label{nsnsnorm}
\ee
Similarly the normalization of the \RR boundary state is
\be
{\cal N}_{\mbox{\scriptsize\RR}}(\hu)=
\sqrt{\hu}{\cal N}_{\mbox{\scriptsize\RR}}=\sqrt{\hu}\frac{\mu_8}{\sqrt{2\pi}}\,,\label{rrnorm}
\ee
$\mu_8$ being the charge density of the BPS D8-brane of Type IIA. Here, we have absorbed the 
factor of $\sqrt{\hu}$ from equation~(\ref{zeroactrr}) into the normalisation
of the \RR sector boundary state for convenience. For 
$\hu=0$ the \RR sector boundary state is zero while 
in the $\hu\rightarrow\infty$ limit we reproduce the usual BPS D8-brane \RR boundary state.
As a check we note that
\be
\bra{0}\left.B,\hu,\eta\right>_{\mbox{\scriptsize\NSNS}}\,,
\ee
reproduce the disc partition function computed in~\cite{GS,KMM1,KMM2,WittenBSFT}. 

As a corollary to the above construction of the \RR sector boundary state for a non-BPS D-brane 
in 
the presence of a linear tachyon it is straightforward to generalise the scattering amplitudes
of~\cite{S1} (see also~\cite{MSS}) 
to obtain the non-BPS D9-brane WZ couplings, including the gravitational piece
\be
S_{\mbox{\scriptsize WZ}}=\frac{\mu_8}{2\sqrt{\pi}}\int C\!\!\w dT\!\! \w\Tr e^F\!\!\w
\sqrt{{\hat A(R)}}e^{-1/4T^2}\,.
\ee
These are in agreement with the results of~\cite{BCR,KMM2}.

\subsection{The cylinder channel}

Having constructed normalised boundary states representing D-branes with a background tachyon 
perturbation, we compute the cylinder diagram corresponding to the exchange of a closed string 
between parallel D-branes. Specifically we are interested in
\be
Z_c(\hu,\hv,l)=\int_0^\infty\rmd l \bra{B,\hv}e^{-lH_{\mbox{\scriptsize c}}}\ket{B,\hu,0}\,,
\ee
where  the bra is computed at 
$\htau=-l$, the ket at $\htau=0$. To match equation~(\ref{bdrbc}) the values of the tachyons are 
\be
\hu=u^2\,,\qquad \hv=-v^2e^{-l}\,.
\ee
In the \NSNS sector the partition function is
\ba
Z_{c,\,\mbox{\scriptsize\NSNS}}(\hu,\hv)
&=&\frac{V_9}{128(2\pi)^{10}}\int_0^\infty\rmd l
\hu(-\hv) 4^{\hu-\hv}B(\hu,\hu)B(-\hv,-\hv)
(\hu-\hv-l\hu\hv)^{-1/2}\nonumber\\& &\qquad\qquad\times
\frac{f_3^7(q)f_3^{(\hu,\hv)}(q)-f_4^7(q)f_4^{(\hu,\hv)}(q)}
{f_1^7(q)f_1^{(\hu,\hv)}(q)}\,,\label{bdrpart}
\ea
and in the \RR sector
\ba
Z_{c,\,\mbox{\scriptsize\RR}}(\hu,\hv)
&=&-\frac{V_9}{64(2\pi)^9}
\int_0^\infty\rmd l \sqrt{-\hu\hv q}
(\hu-\hv-l\hu\hv)^{-1/2}
\frac{f_2^7(q)f_2^{(\hu,\hv)}(q)}{f_1^7(q)f_1^{(\hu,\hv)}(q)}\,.
\ea
where $q=e^{-l}$ and $V_9$ is the (infinite) volume of the directions along which the
D-brane extends apart from $x^9$. The $f_i^{(\hu,\hv)}$ are defined as
\ba
f^{(\hu,\hv)}_1(q)
&=&q^{1/12} \prod_{n=1}^\infty (1-\frac{n-\hu}{n+\hu}\frac{n+\hv}{n-\hv}q^{2n})\,,\nonumber \\
f^{(\hu,\hv)}_2(q)&=&\sqrt{2}\, 
q^{1/12} \prod_{n=1}^\infty (1+\frac{n-\hu}{n+\hu}\frac{n+\hv}{n-\hv}q^{2n})\,,\nonumber\\
f^{(\hu,\hv)}_3(q)&=&q^{-1/24} 
\prod_{r=1/2}^\infty (1+\frac{r-\hu}{r+\hu}\frac{r+\hv}{r-\hv}q^{2r})\,,\nonumber \\
f^{(\hu,\hv)}_4(q)&=&q^{-1/24} 
\prod_{r=1/2}^\infty (1-\frac{r-\hu}{r+\hv}\frac{r+\hv}{r-\hv}q^{2r})\,,
\ea
and $f_i(q)=f^{(0,0)}_i(q)$.
$Z_c$ reproduces the cylinder diagrams in the four conformal limits 
$u,v\rightarrow 0,\infty$, which correspond to NN, ND, DN and DD boundary conditions in the $x^9$ 
direction\footnote{The $\hu,\hv\rightarrow 0$ limit 
is a little more subtle as the Gaussian integral in the direction of the tachyon field now 
becomes part of the volume integral. When evaluating the momentum part of the cylinder 
amplitude we obtained $(\hu-\hv-l\hu\hv)^{-1/2}$,
which is only valid away from the zero tachyon.}. 
The above partition functions are in agreement 
with the ones computed using the Green's function method in Section~\ref{sec2} 
for the case of vanishing gauge fields and one-dimensional tachyons. Due to the closed
string GSO projection these integrals do not have divergences 
corresponding to the closed string tachyon. They do
however, have an open string tachyon divergence signaling an instability of the D9-brane 
vacuum~\cite{Marcus,Bardakci}. Further
there is a divergence due to the massless exchange; it would be interesting to see if this
can be treated using the Fischler-Susskind mechanism~\cite{FishSuss}.

Finally, we have found that the above amplitudes factorise on poles at the on-shell closed string 
mass levels. The residues of these poles are tachyon dependent. 
This suggests that the above partition functions may be interpreted as transition amplitudes for on-shell 
closed string states between D-branes with turned on tachyons.\footnote{We would like to 
acknowledge a discussion with C. Schweigert on this subject.} 

\section{Open string in the presence of a tachyon}\label{sec4}
\setcounter{equation}{0}

In this section we first quantise the open superstring on a strip in the presence of a tachyon; 
we use
these results to compute the one-loop partition function for such a string and identify it with 
the BSFT functional integral on the annulus in the presence of tachyon perturbations. The 
analysis in this section follows the same lines as~\cite{abou}. Consider the action
\be
S=\frac{1}{2\pi}\int_{-\infty}^\infty\rmd\tau\int_0^\pi d\sigma\left(
\partial_\tau X \partial_\tau X-\partial_\sigma X\partial_\sigma X 
\right) 
-\frac{1}{2\pi}\int_{-\infty}^\infty 
\rmd\tau\left[ \uo X^2(\sigma=0)+\vo X^2(\sigma=\pi)\right] \,.
\ee
The constants $\uo\,,\vo$ are related to the tachyons on the annulus via 
(see equation~(\ref{openbc}))
\be
\uo=\frac{u_9^2l}{\pi}\,,\qquad\vo=\frac{e^{-l}v_9^2l}{\pi}\,.
\ee
Varying the action one obtains the usual wave equation 
\be
(\partial^2_\tau-\partial^2_\sigma)X=0\,,
\ee
with boundary conditions
\ba
\partial_\sigma X=\uo X&&\mbox{at $\sigma=0$} \,,\nonumber \\
\partial_\sigma X=-\vo X&&\mbox{at $\sigma=\pi$}\,. \label{openbdrcond}
\ea
The solution is
\be
X=i\sum_{n\neq 0}\alpha_{\epsilon_n}\chi_{\epsilon_n}(\tau,\sigma)\,,
\ee
where 
\be
\chi_{\epsilon_n}=\frac{|c_n|}{\epsilon_n}
\cos\left[\epsilon_n\sigma-\tan^{-1}(\uo/\epsilon_n)\right]
e^{-i\epsilon_n\tau}\,,
\ee
and $\epsilon_n$ is the $n$-th root of the equation
\be
e^{2i(\tan^{-1}(\uo/\epsilon_n)+\tan^{-1}(\vo/\epsilon_n))}=e^{2\pi i\epsilon_n}\,.
\ee
There is a countably infinite number of such solutions satisfying 
$\epsilon_{-n}=-\epsilon_n$. In the above the normalisation constant $c_n=c_{-n}$ is 
\be
\frac{1}{c_n^2}=\frac{\uo+\vo}{\pi}\frac{\epsilon_n^2+\uo\vo}{(\uo^2+\epsilon_n^2)
(\vo^2+\epsilon_n^2)}+1\,.
\ee
The mode functions then satisfy the orthogonality relation 
\be
\int_0^\pi\frac{d\sigma}{\pi}{\bar\chi}_{\epsilon_m}(\tau,\sigma)
(i\stackrel{\leftrightarrow}{\p_{\tau}})\chi_{\epsilon_n}(\tau,\sigma)=\frac{1}{|\epsilon_n|}\d_{mn}\,,
\ee
where $\phi\stackrel{\leftrightarrow}\partial_\tau\psi\equiv
\phi\partial_\tau\psi-\psi\partial_\tau\phi$. 
The canonical momentum $P(\tau,\sigma)$ is defined in the usual way
\be
P(\tau,\sigma)=\frac{\partial{\cal L}}{\partial (\partial_\tau X)} =
\frac{1}{\pi}\partial_\tau X(\tau,\sigma)\,.
\ee
Inverting the expression for $X$ we have 
\be
\alpha_{\epsilon_n}=\epsilon_n\int_0^\pi \frac{d\sigma}{\pi}{\bar\chi}_{\epsilon_n}
(P+\frac{i}{\pi}\epsilon_n X)\,,
\ee
and given the canonical commutation relations 
\ba
\left[X(\tau,\sigma),X(\tau,\sigma^\prime)\right]=0\,,\qquad
\left[P(\tau,\sigma),P(\tau,\sigma^\prime)\right]=0\,,\qquad
\left[X(\tau,\sigma),P(\tau,\sigma^\prime)\right]=
i\delta(\sigma-\sigma^\prime)\,,
\ea
we find that the Fourier modes satisfy the commutation relations
\be
\left[\alpha_{\epsilon_n},\alpha_{\epsilon_m}\right]=\epsilon_m\delta_{m,-n}\,.
\ee
The Hamiltonian is
\ba
H^{\mbox{\scriptsize bos}}_o&=&
\frac{1}{2\pi}\int^\pi_0d\sigma(\partial_\tau^2+\partial_\sigma^2)X(\tau,\sigma)+
\uo X^2(\tau,\sigma)\delta(\sigma)+\vo X^2(\tau,\sigma)\delta(\sigma-\pi)\nonumber \\
&=&\half\sum_{n=1}^\infty\alpha_{\epsilon_{-n}}\alpha_{\epsilon_n}+c(\uo,\vo)\,.
\ea 
In the above $c(\uo,\vo)$ is the normal ordering constant written formally as
\be
c(\uo,\vo)=\frac{1}{2}\sum_{n=1}^\infty \epsilon_n\,,
\ee
which needs to be regularised. A way to compute the regularised $c$ was recently
suggested in~\cite{Bardakci} 
for the case when the tachyons on the two boundaries are the same. Below we generalise slightly 
this computation for the case of distinct tachyons. Consider 
\be
\phi(z)=e^{i\pi z}\frac{z-i\uo}{z+i\uo}-e^{-i\pi z}\frac{z+i\vo}{z-i\vo}\,.
\ee
This function has zeros at $z=\epsilon_n$ and is well defined for all values of the 
tachyons except at the poles $z=-i\uo\,,i\vo$.\footnote{$\phi$ is different from the one used 
in~\cite{Bardakci}. However as we will see, this does not 
change the answer for the integral $I$.} Define
\be
I=\frac{1}{4\pi i}\oint ze^{-\delta z}\rmd\ln\phi\,,
\ee
where the contour encloses the positive real line and therefore the integral is 
equal to $c(\uo,\vo)$ when the regularisation parameter (chosen to have an imaginary part)
$\delta\rightarrow 0$. Now
we open up the contour making it run along the imaginary axis avoiding the two poles at 
$z=-i\uo\,,i\vo$ as in Figure 1. 
\begin{figure}[htb]
\epsfysize=5cm
\centerline{\epsffile{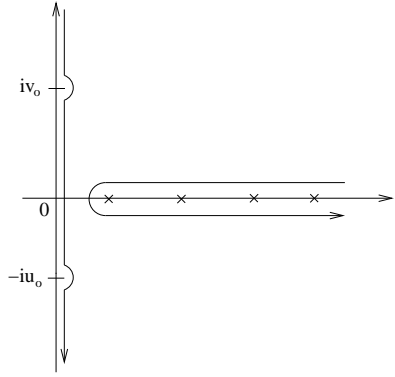}}
\caption{Change of the integration contour for $I$. The $\epsilon_i$ are denoted by crosses.}
\end{figure}
\noindent The integral reduces to
\ba
I&=&\frac{1}{2\pi}\int_0^\infty 
\ln\left(1-e^{-2\pi x}\frac{x-\uo}{x+\uo}\frac{x-\vo}{x+\vo}\right)
\rmd(x\cos(\delta x))
-\frac{1}{2}\int_0^\infty x\cos(\delta x)\rmd x\nonumber \\ &&
-\,\,\frac{1}{4\pi}\int_0^\infty xe^{-i\delta x}
\left(\frac{1}{x+\vo}+\frac{1}{x+\uo}\right)\rmd x
+J(\uo)+J(\vo)\,,
\ea 
where we have separated out the terms $J(\uo)\,,J(\vo)$ 
of $\phi$ which have poles on the imaginary axis and are defined as 
\be
J(\vo)=-\frac{i}{4\pi}\int_C\frac{ze^{-\delta z}}{z-i\vo}\rmd z\,.
\ee
Here $C$ is a contour consisting of three parts: 
$0\le z\le i(\vo-\epsilon)$ for $C_1$, $C_2$ is a small semi-circle 
of radius $\epsilon$ around $i\vo$ and for $C_3$, $i(v+\epsilon)\le z\le \infty$.
In the limit $\epsilon\rightarrow 0$ we obtain
\be
J(\vo)=-\frac{i}{4\pi\delta}-\frac{\vo}{4\pi} e^{-i\delta\vo}\mbox{Ei}(i\delta\vo)\,,
\ee
where $\mbox{Ei}(z)$ is the exponential integral function. The original integral now becomes
\ba
I&=&\frac{1}{2\pi}\int_0^\infty
\ln\left(1-e^{-2\pi x}\frac{x-\uo}{x+\uo}\frac{x-\vo}{x+\vo}\right)
\rmd(x\cos(\delta x))
\nonumber \\
&&+\frac{1}{2\delta^2}+
\frac{1}{4\pi}\left(\frac{i}{\delta}+\vo e^{i\delta \vo}\Gamma(0,i\delta \vo)\right)
-\frac{1}{4\pi}\left(\frac{i}{\delta}+\vo e^{-i\delta \vo}
\mbox{Ei}(i\delta \vo)\right)\nonumber \\
&&
+\frac{1}{4\pi}\left(\frac{i}{\delta}+\uo e^{i\delta \uo}\Gamma(0,i\delta \uo)\right)
-\frac{1}{4\pi}\left(\frac{i}{\delta}+\uo e^{-i\delta \uo}
\mbox{Ei}(i\delta \uo)\right)\,,
\ea
where $\Gamma(x,y)$ is the incomplete Gamma function.
In the limit $\delta\rightarrow 0$ we find
\ba
I&\rightarrow& \frac{1}{2\delta^2}-\frac{\uo+\vo}{2\pi}\ln(i\delta)
-\frac{1}{2\pi}\left(\gamma(\uo+\vo)+\uo\ln\uo+\vo\ln\vo\right)\nonumber \\
&&+\,\,
\frac{1}{2\pi}\int_0^\infty \rmd x\ln\left(
1-e^{-2\pi x}\frac{x-\uo}{x+\uo}\frac{x-\vo}{x+\vo}\right)\,.
\ea
For $\vo=\uo$ this expression agrees with the one obtained in~\cite{Bardakci}.
%\footnote{By adding
%constant boundary tachyonic profiles $T=a_1\,,a_2$ the 
%logarithmic divergence may be removed by renormalising $a_1\,,a_2$, in a fashion similar to the
%closed string channel.}  
The regularised normal ordering constant is
\ba
c(\uo,\vo)&=&\frac{1}{2\pi}\int_0^\infty\rmd x\ln\left(
1-e^{-2\pi x}\frac{x-\uo}{x+\uo}\frac{x-\vo}{x+\vo}\right)
\nonumber \\&&
-\,\,\frac{1}{2\pi}\left(\gamma(\uo+\vo)+\uo\ln\uo+\vo\ln\vo\right)
\,.
\ea
The integral above reproduces the NN and DD normal ordering constants ($-\frac{1}{24}$) corresponding 
to $\uo=\vo=0\,,\infty$, respectively. Further, it also 
reproduces the normal ordering constant of an ND string 
($\frac{1}{48}$) obtained by taking $\uo=0\,, \vo=\infty$. The terms divergent when the tachyons
go to infinity will cancel with terms coming from the fermion normal ordering constant.\footnote{In the
bosonic case these terms diverge and should match the corresponding divergences of the partition
function in the closed string channel. We thank A. Konechny for a discussion on this point.}

Returning to the computation of the annulus diagram in the open string channel it is not 
difficult to compute the partition function for a single boson with boundary 
conditions~(\ref{openbdrcond})
\be
Z_{\mbox{\scriptsize bosonic}}(\uo,\vo)=
\Tr(e^{-2\pi tH^{\mbox{\scriptsize bos}}_o})={\tilde q}^{2c(\uo,\vo)}
\prod_{n=1}^\infty (1-{\tilde q}^{2\epsilon_n})^{-1}\,,\label{bpart}
\ee
with ${\tilde q}=e^{-\pi t}$ and $t=\pi/l$.

A similar analysis has been carried out for the fermions. 
In the R sector these have the same 
moding as the bosons; canonically quantised they satisfy the anti-commutation relations
\be
\{\psi_{\epsilon_n},\psi_{\epsilon_n}\}=\delta_{m,-n}\,,
\ee
and have the Hamiltonian
\be
H^{\mbox{\scriptsize R}}_{\mbox{\scriptsize o}}
=\frac{1}{2}\sum_{n=1}^\infty\psi_{\epsilon_{-n}}\psi_{\epsilon_n}
-c(\uo,\vo)\,.
\ee
In the NS sector the moding is different with the $\epsilon_r$ now satisfying
\be
e^{2i(\tan^{-1}(\uo/\epsilon_r)+\tan^{-1}(\vo/\epsilon_r))}=-e^{2\pi i\epsilon_r}\,.
\ee
The anti-commutation relations and Hamiltonian are
\be
\{\psi_{\epsilon_r},\psi_{\epsilon_s}\}=\delta_{r,-s}\,,\qquad 
H^{\mbox{\scriptsize NS}}_{\mbox{\scriptsize o}}
=\frac{1}{2}\sum_{r=1}^\infty\psi_{\epsilon_{-r}}\psi_{\epsilon_r}
-c_{\mbox{\scriptsize NS}}(\uo,\vo)\,,
\ee
where in the NS sector the regularised normal ordering constant is 
\ba
c_{\mbox{\scriptsize NS}}(\uo,\vo)&=&\frac{1}{2\pi}\int_0^\infty dx\ln\left(
1+e^{-2\pi x}\frac{x-\uo}{x+\uo}\frac{x-\vo}{x+\vo}\right)\nonumber \\ &&
-\,\,\frac{1}{2\pi}\left(\gamma(\uo+\vo)+\uo\ln\uo+\vo\ln\vo\right)
\,.
\ea
The partition function for a NS fermion is
\be
Z_{\mbox{\scriptsize NS}}(\uo,\vo)=
\Tr_{\mbox{\scriptsize NS}}(e^{-2\pi tH^{\mbox{\scriptsize NS}}_o})=
{\tilde q}^{-2c_{\mbox{\scriptsize NS}}(\uo,\vo)}
\prod_{r>0}^\infty(1+{\tilde q}^{2\epsilon_r})\,,\label{nspart}
\ee
while that of an R fermion is
\be
Z_{\mbox{\scriptsize R}}(\uo,\vo)=\Tr_{\mbox{\scriptsize R}}
(e^{-2\pi tH^{\mbox{\scriptsize R}}_o})=
{\tilde q}^{-2c(\uo,\vo)}\prod_{n=1}^\infty 
(1+{\tilde q}^{2\epsilon_n})\,.\label{rpart}
\ee
It is clear by construction and inspection that the coefficients 
of ${\tilde q}$ are integral. This is expected of an open string partition function.

Combining equations~(\ref{bpart}),~(\ref{nspart}) and~(\ref{rpart})
the partition function for open strings on a non-BPS D-brane in the presence of 
background tachyons is
\be
Z_{\mbox{\scriptsize open}}=\int_0^\infty\frac{\rmd t}{2t}\Tr_{\mbox{\scriptsize NS-R}}
(e^{-2\pi tH_{\mbox{\scriptsize o}}})
=\frac{V_9}{(2\pi)^9}\int_0^\infty\frac{\rmd t}{2t}(2t)^{-9/2}
\frac{f_3^7({\tilde q})g^{(\uo,\vo)}_3({\tilde q})-f_2^7({\tilde q})g^{(\uo,\vo)}_2({\tilde q})}
{f_1^7({\tilde q})g^{(\uo,\vo)}_1({\tilde q})}\,,\label{openpart}
\ee
where $V_9$ is the volume of space-time with no background tachyon, and the $g$ functions are defined
as 
\ba
g^{(\uo,\vo)}_1({\tilde q})&=&{\tilde q}^{-2c(\uo,\vo)}
\prod_{r>0}^\infty(1-{\tilde q}^{2\epsilon_n})\,,\qquad
g^{(\uo,\vo)}_2({\tilde q})={\tilde q}^{-2c(\uo,\vo)}
\prod_{r>0}^\infty(1+{\tilde q}^{2\epsilon_n})\,, \\
g^{(\uo,\vo)}_3({\tilde q})&=&{\tilde q}^{-2c_{\mbox{\scriptsize NS}}(\uo,\vo)}
\prod_{r>0}^\infty(1+{\tilde q}^{2\epsilon_r})\,,\qquad
g^{(\uo,\vo)}_4({\tilde q})={\tilde q}^{-2c_{\mbox{\scriptsize NS}}(\uo,\vo)}
\prod_{r>0}^\infty(1-{\tilde q}^{2\epsilon_r})\,.
\ea
It is easy to see that in equation~(\ref{openpart}) the terms divergent for 
$\uo\,,\vo\rightarrow\infty$ in the
normal ordering constants of bosons and fermions cancel.

In this section we have computed  the partition function on the annulus by an operator method, slicing
time  in the periodic direction. In the previous section we used a different operator formalism with
time running from one boundary of the annulus to the other. Since both approaches compute the same
quantity we expect that equations~(\ref{bdrpart}) and~(\ref{openpart}) should give the same result. 
For conformal theories this is easily checked using the $t\rightarrow l$ 
transformation properties of the $f_i$ 
functions. Unfortunately we do not know the corresponding transformations for the $g^{(\uo,\vo)}_i$ and 
$f^{(\hu,\hv)}_i$ functions and are, as a result, unable to verify this claim directly.

In the closed string channel discussed in Section~\ref{sec3} the \RR partition function 
gave a non-zero answer (for $\hu\,,\hv\neq 0$). This can be re-interpreted as the statement 
that there is a GSO-like projection acting on open strings in the presence of non-zero tachyons 
on the
boundary. This projection should presumably be defined as a mod 2 number operator for 
world-sheet fermions just as in the case without background tachyon. This suggests a further 
contribution
to the partition function of interest of the form
\be
Z_{\mbox{\scriptsize open}}=\int_0^\infty\frac{\rmd t}{2t}\Tr_{\mbox{\scriptsize NS-R}}
(e^{-2\pi tH_{\mbox{\scriptsize o}}}(-1)^F)
=-\frac{V_9}{(2\pi)^9}\int_0^\infty\frac{\rmd t}{2t}(2t)^{-9/2}
\frac{f_4^7({\tilde q})g^{(\uo,\vo)}_4({\tilde q})}
{f_1^7({\tilde q})g^{(\uo,\vo)}_1({\tilde q})}\,,
\ee
where due to zero-modes the R sector trace is zero.

\section{Conclusion} 

We have computed the partition function in the
supersymmetric boundary sigma model on the annulus for the case of the
exactly solvable linear tachyon profile. We showed how the one and the
same answer can be achieved by means of different techniques: the Green's
function method and the boundary state formalism, justifying thereby the
latter for the case of non-conformal boundary deformations. An interesting
feature of our results is that the \RR sector provides a non-trivial
contribution to the partition function. If the 
interpretation of the annulus partition function as a one-loop correction
to the space-time effective action $S_{\mbox{\scriptsize eff}}$ is correct one may determine the
corresponding change in $S_{\mbox{\scriptsize eff}}$. Taking the tachyon profile to  be the same on the 
two boundaries one may expand the partition function around small $u$ and interpret it as coming from
an effective space-time action for the tachyon field $T$. In the \RR sector only zero-modes
contribute to the leading order in $u$-expansion and one finds
\be
S_{\mbox{\scriptsize eff, \RR}}^{1-loop}\sim 
\int \rmd^{10}x\int_0^1\frac{\rmd a}{a}K(a)
e^{-\frac{1}{4}(1+a)T^2}[\partial_{\mu}T\partial^{\mu}T+F^2_{\mu\nu}\partial^\mu T\partial^\nu T
-\frac{1}{2}\partial_\mu T\partial^\mu T F_{\nu\rho}F^{\nu\rho}+\dots]\,,\label{rract}
\ee
where we have indicated by dots the higher derivative terms and the mixing between the tachyon 
and the gauge field comes from the zero-modes as discussed in Section~\ref{sec33}. In the
above 
\be
K(a)=\frac{f_2^8(a)}{f_1^8(a)}\,.
\ee
Thus, the \RR sector
contributes only to the derivative terms and not to the tree
level potential. Similarly expanding the \NSNS partition function in $u$ one finds its contribution
to the effective action as
\be
S_{\mbox{\scriptsize eff, \NSNS}}^{1-loop}\sim 
\int d^{10}x\int_0^1\frac{da}{a}K(a)
e^{-\frac{1}{4}(1+a)T^2}[1+b(a)\partial_{\mu}T\partial^{\mu}T+\dots]\,.\label{nsnsact}
\ee
Here $b(a)$ is the next-to-leading term in the $u$-expansion of the \NSNS partition function.

The integral over $a$ diverges due to the $a\rightarrow 1$ behaviour of the integrand.
This arises from the open string tachyon and should be subtracted in a manner compatible with 
open/closed string duality. It is desirable to find such a subtraction scheme. This divergence indicates
the instability of the $T=0$ vacuum~\cite{Marcus,Bardakci}.
In obtaining the partition function in the closed string channel we have summed over the
spin structures, thus removing the closed string tachyon. This manifests itself in the fact that
the $a$ integral above has no linear divergences for $a\rightarrow 0$. However, there is
a logarithmic divergence in the $a\rightarrow 0$ limit corresponding to a massless exchange. 
It is an interesting question whether this divergence (in the case of the superstring) can be
treated using a Fischler-Susskind type mechanism~\cite{FishSuss}.

We have also discussed the canonical
quantization of the open string in the presence of a linear tachyon
background. An important issue here is a generalization of the open/closed
duality for this background. To compute the partition function in the open
string channel one should know the normal ordering
constant of the open string Hamiltonian that depends on the tachyon profile in
a non-trivial way. Clearly the normal ordering constant is divergent and usually one
picks up a subtraction scheme to define a finite quantity $c$. We discussed such a scheme 
(generalising the one in~\cite{Bardakci}) accounting for two different 
boundary tachyons. The part of $c$ that is finite when $u$ or $v$ goes to infinity reproduces
the normal ordering constants for the NN,DD and ND cases.

\vspace{1cm}
{\bf Note added.} When our work was completed an interesting paper~\cite{CKL} appeared where the
some issues related to the construction of the loop corrected tachyon potential were discussed.

\section*{Acknowledgements} 

We would like to thank M. Bianchi, M.B. Green, I. Runkel, C. Schweigert, P. Vanhove
and particularly S. Silva
for many helpful conversations and useful insights. S. Frolov has provided many
valuable comments on the manuscript.
We are especially grateful to S. Theisen for
initially collaborating on the project, for reading the manuscript and for sharing his insights.
G.A. was supported by the DFG and in part by RFBI grant N99-01-00166 and
by INTAS-99-1782. We are supported by the European Commission RTN programme HPRN-CT-2000-00131 in which we
are associated to U. Bonn. A.P. also acknowledges support from GIF, the
German-Israeli foundation for Scientific Research under grant number I-522-010.07/97. 

\appendix

\section{Derivation of Green's functions and the partition function}
\setcounter{equation}{0}
In the Appendix we present the explicit expressions for the various Green's functions. We 
discuss in some detail the bosonic boundary conditions, their relation to Gauss's theorem 
and the transposition properties of the Green's function. Finally, we briefly outline how
to obtain the partition function from the Green's function.

\subsection{Green's functions}
\subsubsection{The bosonic sector}
Consider the more general boundary conditions for the bosonic Green's function 
\ba\label{bcb}
(z\p_z+\bz\p_{\bz})G(z,w)+uG(z,w)+F(z\p_z-\bz\p_{\bz})G(z,w)&=&D,\qquad |z|=1\,,\nonumber \\
-(z\p_z+\bz\p_{\bz})G(z,w)+avG(z,w)+L(z\p_z-\bz\p_{\bz})G(z,w)&=&E,\qquad |z|=a\,,
\ea
where $D$ and $E$ are yet unknown matrices that may depend on the fields. 
Since $G_{\m\n}(z,w)=\la X_{\m}(z)X_{\n}(w)\ra$ the Green's function must satisfy
\begin{equation}\label{sym}
G_{\m\n}(z,w)=G_{\n\m}(w,z).
\end{equation}
  
To find the Green's function corresponding to the boundary conditions~(\ref{bcb}) we make the ansatz 
\begin{equation}
G(z,w)=G_f(z,w)+A+B\ln|z|^2\ln|w|^2+C\ln|z|^2+C^T\ln|w|^2+\sum_{k\in\Z\setminus\{0\}}
\bigl(\a_{k}(w)z^k+\bar{\a}_{k}(w)\bz^k\bigr),
\end{equation}
where $G_f(z,w)=-\ln|z-w|^2$ is the fiducial Green's function, obeying ($\a'=2$) 
\begin{equation}
-\frac{1}{2\pi}\p_z\p_{\bz}G_f(z,w)=\d^{(2)}(z,w).
\end{equation}
Moreover we require that $A$, $B$ and $C$ are real matrices, satisfying
\begin{equation}\label{sym1}
A=A^T,\qquad B=B^T.
\end{equation}
  
The derivation of the oscillators is straightforward and one finds
\ba
\a_k&=&\frac{1}{k}\left(1-a^{2k}S_k(u,F)S_k(av,L)\right)^{-1}
S_k(u,F)\left(\bw^k+a^{2k}w^{-k}S_k(av,L)\right)\,,\nonumber \\
\a_{-k}&=&\frac{a^{2k}}{k}\left(1-a^{2k}S_k^T(av,L)S_k^T(u,F)
\right)^{-1}S_k^T(av,L)\left(\bw^{-k}+w^kS_k^T(u,F)\right)\,,\nonumber \\
\tilde{\a}_k&=&\frac{1}{k}\left(1-a^{2k}S^T_k(u,F)S_k^T(av,L)\right)^{-1}S_k^T(u,F)
\left(w^k+a^{2k}\bw^{-k}S_k^T(av,L)\right)\,,\nonumber \\
\tilde{\a}_{-k}&=&\frac{a^{2k}}{k}\left(1-a^{2k}S_k(av,L)
S_k(u,F)\right)^{-1}S_k(av,L)\left(w^{-k}+\bw^kS_k(u,F)\right).
\ea
It is easy to see that the oscillator dependent parts of the Green's function 
indeed satisfy~(\ref{sym}). For the zero-modes we find the conditions
\ba
\label{0a}-2+2C+uA-D+\bigl(2B+uC^T\bigr)\ln|w|^2&=0\,,\quad |z|=1\,,\\
\label{0b}-2C+av(A+2C\ln a)-E+\bigl(-2B-av+2avB\ln a +avC^T\bigr)\ln|w|^2&=0\,,\quad |z|=a\,.
\ea
Taking into account~(\ref{sym1}), the cancellation of the $\ln|w|^2$ dependent 
terms in~(\ref{0a}) and~(\ref{0b}) requires that
\begin{equation}
B=-\frac{1}{2}Cu\quad\mbox{and}\quad uC^T=Cu
\end{equation}
and $C$ is explicitly found to be
\begin{equation}
C=av(u+av-auv\ln a)^{-1}.
\end{equation}
From this expression one may check that $C^T$ indeed satisfies the constraint $uC^T=Cu$. 
Solving~(\ref{0a}) one finds
\begin{equation}
uA=\bigl(2(1-C)+D\bigr)\,,\qquad uD^T=Du
\end{equation}
and~(\ref{0b}) implies
\begin{equation}\label{de}
Eu=avD^T\,.
\end{equation}
Thus, the matrices $D$ and $E$ are related by~(\ref{de}) but not completely 
fixed by the boundary conditions. Since a 
non-vanishing $D$ only results in a change of the overall normalisation of the 
(bosonic) partition function it is 
convenient to choose $D=0=E$. Then $A$ is explicitly given by 
\begin{equation}
A=2(1-av\ln a)(u+av-auv\ln a)^{-1},
\end{equation}
modulo elements in the kernel of $u$ which we suppress for the same reasons as a non-vanishing $D$.

As a final check, we prove that Gauss's theorem is satisfied for any $D$. 
Since the Green's function satisfies
\begin{equation}
\Box G(\s_1,\s_2)=-4\pi\d^{(2)}(\s_1,\s_2)
\end{equation}
together with the boundary conditions~(\ref{bcb}), Gauss's theorem requires
\begin{equation}
-4\pi=\int_{\p\S}\p_rGds=-u\int_{|z|=1}d\phi\,G(z,w)-av\int_{|z|=a}d\phi\,G(z,w)+2\pi(D+E)\,.
\end{equation}
Only the integrals over the zero-modes will be non-vanishing and, therefore 
we confirm that
\begin{equation}
0=4\pi-2\pi\bigl(uA+avA+2avC\ln a-D-E+\ln|w|^2(uC^T-av+2avB\ln a+avC^T)\bigr)\,.
\end{equation}

\subsubsection{The \NSNS sector}

Here we present the explicit expressions for the oscillators of the various Green's functions in the
\NSNS sector. For example, the $G_{++}$ oscillators are
\ba
\psi_r(w)&=&i\eta\tilde{\eta}a^{2r}\left(1-\eta\tilde{\eta}a^{2r}S_r(u,F)
S_r(av,L)\right)^{-1}S_r(u,F)S_r(av,L)w^{-r}\,,\nonumber \\
\psi_{-r}(w)&=&-i\eta\tilde{\eta}a^{2r}\left(1-\eta\tilde{\eta}a^{2r}S_r^T(av,L)S_r^T(u,F)
\right)^{-1}S_r^T(av,L)S_r^T(u,F)w^r\,.
\ea
Similarly 
\ba
\tilde{\psi}_r(\bw)&=&i\eta\tilde{\eta}a^{2r}\left(1-\eta\tilde{\eta}a^{2r}
S_r^T(u,F)S_r^T(av,L)\right)^{-1}S_r^T(u,F)S_r^T(av,L)\bw^{-r}\,,\nonumber \\
\tilde{\psi}_{-r}(\bw)&=&-i\eta\tilde{\eta}a^{2r}\left(1-\eta\tilde{\eta}a^{2r}S_r(av,L)
S_r(u,F)\right)^{-1}S_r(av,L)S_r(u,F)\bw^r\,,\\
a_r(\bw)&=&\eta\left(1-\eta\tilde{\eta}a^{2r}S_r(u,F)S_r(av,L)\right)^{-1}S_r(u,F)\bw^r\,,
\nonumber \\
a_{-r}(\bw)&=&-\tilde{\eta}a^{2r}\left(1-\eta\tilde{\eta}a^{2r}
S_r^T(av,L)S_r^T(u,F)\right)^{-1}S_r^T(av,L)\bw^{-r}\,,
\ea
and
\ba
b_r(w)&=&-\eta\left(1-\eta\tilde{\eta}a^{2r}S_r^T(u,F)
S_r^T(av,L)\right)^{-1}S_r^T(u,F)w^r\,,\nonumber\\
b_{-r}(w)&=&\tilde{\eta}a^{2r}\left(1-\eta\tilde{\eta}a^{2r}S_r(av,L)
S_r(u,F)\right)^{-1}S_r(av,L)w^{-r}\,.
\ea

\subsubsection{The \RR sector}

For the non-zero modes the oscillators are exactly the same as in the \NSNS sector, 
the only difference being that now $r$ is an integer.
  
\subsection{The partition function}
In this sub-section we outline the derivation of the partition function in the closed channel 
following a technique used in~\cite{WittenBSFT}. 
For simplicity we restrict ourselves to the bosonic contribution, the derivation of the 
fermionic parts proceeds exactly along the same lines.

Differentiation of the (bosonic part of the) boundary action with respect to
the couplings $u$ and $v$ results in the differential equations 
\ba
\frac{\p\ln Z}{\p u^{\m\n}}&=&-\frac{1}{8\pi}\int_0^{2\pi}d\phi\la X_{\m}(\phi)X_{\n}(\phi)\ra=
-\frac{1}{8\pi}\int_0^{2\pi}d\phi\,G_{\m\n}(e^{i\phi},e^{i\phi})\,,\nonumber\\
\frac{\p\ln Z}{\p v^{\m\n}}&=&-\frac{1}{8\pi}\int_0^{2\pi}d\phi\,aG_{\m\n}(ae^{i\phi},ae^{i\phi})
\ea
and similar equations when differentiating with respect to $F$ and $L$. Using the result 
for the bosonic Green's function it is 
not difficult to verify that the solution to these equations (up to the normalization) is
\ba
Z_{\mbox{\scriptsize bos}}&=&\det(u+av-auv\ln a)^{-1/2}
\prod_{k=1}^{\infty}\det(1+u/k+F)^{-1}\det(1+av/k+L)^{-1}\nonumber\\&&\times\,\,
\det\left(1-a^{2k}S_k(u,F)S_k(v,L)\right)^{-1}\,.
\ea
$Z$ also satisfies the equations obtained by differentiating 
with respect to $F_{\m\n}$ and $L_{\m\n}$. In 
principle there may be an overall dependence on the modulus $a$ in the partition 
function that could not be fixed by the 
previous considerations. However, one can also derive an equation for 
$\frac{\p\ln Z}{\p a}$ by looking at the change 
of the action under variations of the modulus~\cite{abou}. Following~\cite{abou} this equation reads
\ba
\p_a\ln Z&=&-\la\p_aS_{\mbox{\scriptsize bulk}}\ra-\la\p_aS_{\mbox{\scriptsize bndy}}\ra\nonumber\\
&=&\frac{1}{2\pi}\frac{a}{1-a^2}\int_{\S}d^2z\,\left(\frac{1}{\bz^2}\la T_{zz}\ra
+\frac{1}{z^2}\la T_{\bz\bz}\ra\right)
-\frac{1}{8\pi}\int_0^{2\pi}d\phi\,v_{\m\n}G^{\m\n}(ae^{i\phi},ae^{i\phi})\,.
\ea
Using the explicit expression for the bosonic Green's function and 
\begin{equation}
\la T_{zz}(z)\ra=-\frac{1}{2}\lim_{w\to z}\left[\p_z\p_wG^{\m}_{\m}(z,w)+\frac{10}{(z-w)^2}\right]
\end{equation}
and similarly for $\la T_{\bz\bz}\ra$ we find that $\p_a\ln Z$ integrates to~(\ref{Zb}), 
so that there is no further dependence on the modulus.

\ed